\documentclass[10pt, reprint, aps, amsmath,amssymb,twocolumn,article]{revtex4-1}

\usepackage{epigraph}
\usepackage{graphicx}
\usepackage{dcolumn}
\usepackage{algorithm,algpseudocode}
\usepackage{bm}
\usepackage[font=scriptsize]{caption}
\usepackage{booktabs}
\usepackage{subcaption,graphicx}
\algnewcommand{\To}{\textbf{To }}
\algnewcommand\Input{\item[\textbf{Input:}]}%
\algnewcommand\Output{\item[\textbf{Output:}]}%

\begin{document}

\title{On the Effects of Resistive and Reactive Loads on Signal Amplification }

\author{Luciano da F. Costa}
\email{luciano@ifsc.usp.br}
\affiliation{S\~ao Carlos Institute of Physics, IFSC-USP,  S\~ao~Carlos, SP,~Brazil}

\date{\today}

\begin{abstract}
The effects of reactive loads into amplification is studied. A simplified common emitter circuit configuration was adopted and respective time-independent and time-dependent voltage and current equations were obtained. As phasor analysis cannot be used because of the non-linearity, the voltage at the capacitor was represented in terms of the respective integral, implying a numerical approach. The effect of purely resistive loads was investigated first, and it was shown that the fanned structure of the transistor isolines can severely distort the amplification, especially for $V_a$ small and $s$ large. The total harmonic distortion was found not to depend on $V_a$, being determined by $s$ and the load resistance $R$. An expression was obtained for the current gain in terms of the base current and it was shown that it decreases in an almost perfectly linearly fashion with $I_B$. Remarkably, no gain variation, and hence perfectly linear amplification, is obtained when $R=0$, provided maximum power dissipation limits are not exceeded. Capacitive loads imply the detachment of the circuit trajectory from a straight line to an ``ellipsoidal''-like loop. This implies a gain asymmetry along upper or lower arcs of this loop. By using the time-dependent circuit equations, it was possible to show numerically and by an analytical approximation that, at least for the adopted circuit and parameter values, the asymmetry induced by capacitive loads is not substantial. However, capacitive loads will imply lag between the output voltage and current and, hence, low-pass filtering. It was shown that smaller $V_a$ and larger $s$ can substantially reduce the phase lag, but at the cost of severe distortion.    
\end{abstract}

\keywords{Amplifiers, Early model, equivalent circuits, gain, total harmonic distortion, linearity, reactive loads.}
\maketitle

\setlength{\epigraphwidth}{.49\textwidth}
\epigraph{\emph{``There is geometry in the humming of the strings.''}}{Pythagoras}

\section{Introduction}

The dynamics of the natural world is underlain by myriad signals covering all scales of \emph{magnitude}.  From the
light of the faintest star to the sound of the loudest thunder (not to mention quarks and supernovae), signals permeate and
integrate the whole universe.  From Pitagora's time, it has been known that natural signals (in that case, sound) can be understood 
in terms of basic harmonic functions, each with a respective magnitude.  Interestingly, the way in which signals are changed by
environment, such as the propagation of a bird songs throughout a forest, can be understood as changes, in specific
ways, in the magnitudes of each of the signal harmonic components.  For simplicity's sake, these signal magnitude alterations 
are henceforth called \emph{amplification}, even when the resulting signal has its amplitude reduced.  So, it can be said that a 
good deal of signal processing resumes to magnitude alterations.  It is consequently hardly surprising that the development of science 
and technology has relied so much on the acquisition, amplification and analysis of signals.  Indeed, all instruments critical for measuring
nature, from microscopes to telescope arrays, require accurate, reliable and linear amplification of signals.  A very much similar situation is
found regarding the vast majority of human activity: linear signal amplification is required in telecommunications, health/leisure, and transportation,
among many other critical areas.  Interestingly, sometimes humans also aim for non-linear amplification, as is the case with music
instrument synthesis.  

\emph{Analog} (or \emph{linear}) electronics (e.g.~\cite{analogdesign:1988,analogdesign:1991}), 
more recently accompanied by digital amplification (e.g.~\cite{menchi:2016}), is the main 
scientific-technologic area focusing on the amplification of natural signals.  It has provided the foundation for much of instrumentation
used in science and technology.  The main goal here often is to achieve as much linear operation as possible, which is a rather challenging
issue given the intrinsically non-linear nature of real-world components, such as transistors.  Interestingly, our world is predominantly
non-linear, being rather difficult to think of natural phenomena that are perfectly linear.  At the same time, most of our approaches to science,
at least until recently, relied on mapping non-linear real-world phenomena into approximated linear representations.  This also happened
with transistor studies, as often (but not always) these devices have been mapped into linear respective representations.  

Indeed, because of its many important and critical applications, much effort has been invested in deriving effective approaches to transistor
characterization, representation and modeling (e.g.~\cite{stewart:1956,jaeger:1997,sedra:1998,horowitz:2015,streetman:2016}).  
Though in the beginning, especially from the 50's to the 70's, these efforts tended to focus 
on graphic and analytical approaches --- with extensive use of characteristic curves, surfaces, and transfer functions 
(e.g.~\cite{shea:1955,zimmermann:1959,pettit:1961,fontaine:1963,alley:1966,gronner:1970,tinnell:1972}), the progressive extension to 
larger and more complex integrated devices was largely assisted by more systematic application of numeric computational simulations.  
Yet, continuing research into more analytical approaches to transistor characterization, representation and modeling remains an important
issue as it can not only extend our intuitive understanding of transistor operation, but also contribute do more effective and accurate
simulation approaches and tools.  Recently, a simple transistor model, as well as its respective equivalent circuit, was 
developed~\cite{costaearly:2017,costaearly:2018,costaequiv:2018} 
that is founded on the Early effect, discovered by J. M. Early in 1952~\cite{early:1952,ziel:1968,streetman:2016}.  
Despite its great simplicity, consisting of a Thevenin configuration
with a fixed voltage source $V_a$ in series with a variable resistance $R_o(I_B)= 1/tan(s I_B)$ ($I_B$ is the current base, and $s$ a
proportionality parameters), this model exhibits two interesting and important features: (i) it allows the representation of the transistor 
varying output resistance in terms of the modulating base current  (corresponding to the fanned isolines that are so characteristic real-world 
transistor graphic representations); and (ii) the two involved parameters, namely $V_a$ and $s$ \emph{do not vary} with the transistor 
operation as quantified by the collector current $I$ and voltage $V_C$.  Recall that both the parameters $\beta$ and $R_o$ adopted in many
traditional transistor modeling approaches are explicit functions of $I$ and $V_C$, i.e.~$\beta = \beta(I,V_C)$ and $R_o = R_o(I,V_C)$.
This parametric dependence implied that averages, or other functionals, of these two parameters had to be used for the characterization
of transistor properties~\cite{costaearly:2018}, implying in specific or arbitrary choices about in which domain to integrate the 
parameters $\beta = \beta(I,V_C)$  and $R_o = R_o(I,V_C)$ into respective averages $\langle \beta \rangle$ and $\langle R_o \rangle$.

Despite its recent introduction, the Early modeling approach has already led to interesting results, such as the characterization of typical
properties of NPN and PNP bipolar junction transistors (BJTs)~\cite{costaearly:2018}, the study of germanium alloy junction 
devices~\cite{germanium:2018},  the derivation of a \emph{prototypic parameter space} characterizing trade-offs between gain and 
linearity~\cite{costaequiv:2018}, the study of stability of circuit operation under voltage supply oscillations~\cite{costaequiv:2018}, as well as the 
derivation of the Early equivalent model of parallel transistor configurations~\cite{costaequiv:2018}.  The aforementioned study of
circuit stability involved a comparison with a more traditional simplified modeling approach, which showed that largely divergent results
can be obtained when considering or not the fanned structure of the base current-indexed isolines, and justifying the use of models such
as the Early approach.  Indeed, the development of an equivalent circuit for the Early model~\cite{costaequiv:2018} paved the way to
applying this modeling approach to a large number of relevant problems regarding transistor characterization, modeling, simulation, and design
in discrete and integrated electronics.  

The present work addresses the issue of transistor amplification in circuits involving reactive loads, a situation very commonly found
in practice.  While purely resistive loads imply a straight line in the transistor characteristic surface, capacitive or inductive loads will
define ``ellipsoidal''-like pathways in these parametric spaces, also implying phase shifts between the current and voltage at the load.  
While these alterations would still lead to linear amplification were the transistor a linear device (i.e.~had parallel, equispaced current 
modulated-isolines), the fanned structure actually observed in real-world devices implies that the achieved amplification will become
non-linear not only because the load line will no longer be straight, but also because different paths will be followed for the positive
and negative portions of the amplified signal. The fact that the induced phase shifts depend on the frequency of the input signal 
complicates further the situation.  The availability of the Early model and respective equivalent circuit provides a viable and
yet accurate way to approach this interesting and important problem of non-linear amplification considering reactive loads, and this
constitutes the main motivation of the  current work.

A simplified common emitter configuration is adopted in which there is no emitter resistance (no negative feedback) and the
load is attached between the transistor collector and the external voltage supply $V_{CC}$.
By using the Early equivalent circuit~\cite{costaequiv:2018}, it was possible to obtain the time-independent and time-dependent
equations for the currents and voltages in the adopted circuit. The case of purely resistive loads is studied first by using 
the respective time-independent circuit equations, which allowed the total harmonic distortion of the amplification to be estimated
for each of the possible configurations in a region of the Early space parameter, leading to remarkable results.  The case of
capacitive loads is studied next, and its effect on linearity and phase lag (low-pass filtering) are studied by using numeric as
well as analytical approximations, also with surprising results.  The work concludes by identifying prospects for future related
research.

\section{A Simplified Common Emitter Configuration and its Early Model Representation}

\begin{figure}[h!]
\centering{
\includegraphics[width=5cm]{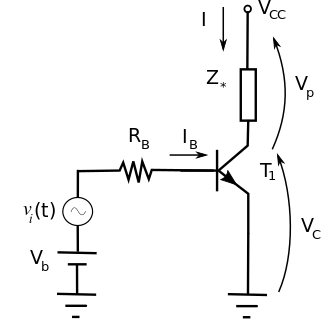}
\caption{The simplified common emitter configuration adopted in the present work.  The 
simplifications do not significantly impact the respective circuit analysis, which can be immediately
extended to other circuit configurations.}
\label{fig:commemitt}}
\end{figure}

Figure~\ref{fig:commemitt} illustrates the simplified common emitter configuration considered in this
work, where $Z_*$ stands for a generic load with resistive and reactive properties.  The adopted
circuit configuration is said to be simplified because the base is does not incorporate some features typically
found in the common emitter configuration, i.e.: (i) the base is not biased by a voltage divider,
and the input signal is driven through a base resistance $R_B$; (ii) there is no resistance $R_E$ 
between the emitter and the ground, which is often used to
achieve negative feedback; and (iii) the load is attached directly between the collector and the 
external voltage supply $V_{CC}$, not attached to the collector through an additional decoupling capacitor. 

The adopted simplifications, however, do not change much
the properties of the circuit with respect to the more traditional common emitter
configuration.  For instance, collector bias can be incorporated as part of the input current $I_B$
so that, mathematically, the same effect can be achieved.  The avoidance of the emitter resistance
is compatible with the objective of the present work, which focus on the characterization of the
circuit linearity under reactive loads, so that it is more interesting to consider the situation with
more distortion.  The situation with $R_E$ will be considered in a forthcoming work focusing on
the study of negative feedback by using the Early formulation.  Regarding simplification (iii),
this does not affect the characterization of the linearity of the transistor and the so-developed 
analysis can be immediately extended to other situations more commonly found in practice,
such as attaching the load to the collector via a decoupling capacitor.

\begin{figure}[h!]
\centering{
\includegraphics[width=8cm]{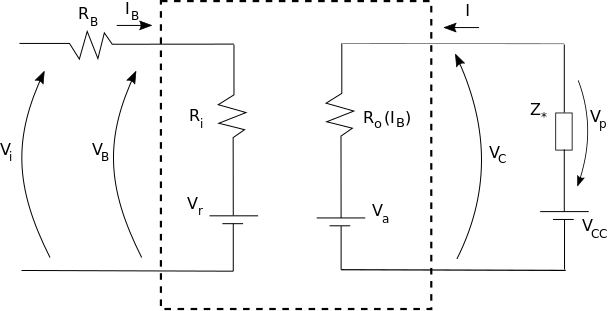}
\caption{The equivalent circuit obtained by replacing the transistor in Figures~\ref{fig:commemitt} by
the respective Early model equivalent circuit.}
\label{fig:equiv}}
\end{figure}

The Early model equivalent circuit~\cite{costaequiv:2018} can be easily employed to obtain the mathematical 
equations underlying the behavior of the circuit in Figure~\ref{fig:commemitt}.  First, we replace the 
transistor by the Early model equivalent circuit~\cite{costaequiv:2018}, yielding the circuit shown in 
Figure~\ref{fig:equiv}.

The input section is modeled in terms of a diode, as usual, so that we have: 

\begin{eqnarray}
   I_B \left( I_B \right) = \frac{V_i - V_r}{R_B + R_i} \\
   V_B \left( I_B \right) =  V_R + \frac{V_i - V_r}{R_B + R_i} R_i   
\end{eqnarray}

Recall that the Early representation of the output port involves two parameters: the Early voltage $V_a$ and a
proportionality parameter $s$.  One of the distinguishing features of the Early model is that
both these parameters do not depend on $I$ or $V_C$.   In addition, these parameters have been
formally related to averages of the more frequently adopted current gain and output resistance.
Interesting, as a consequence of scaling properties of the fanning geometrical structure underlying the
Early model, the effective current gain results, in an approximate sense, directly proportional to the product of
$V_a$ magnitude and $s$, i.e.~ $\langle \beta \rangle \approx s \left| V_a \right|$.  
This can be immediately appreciated by considering that: (a) if $V_a$ is doubled, the effective current 
gain will be halved; (b) if $s$ is doubled, this same gain is doubled.   So, the net product is not to change 
the effective current gain.  We also have that $\langle R_o \rangle \propto \left| V_a \right|$.  
It is interesting to perform the circuit analyses and discussions considering distinct parameter configurations
leading to the same current gain.

In the respective equivalent circuit~\cite{costaequiv:2018},
the parameter $s$ defines the output resistance $R_o (I_B) = 1/tan(s I_B)$ that is a function
of the base current $I_B$ given a fixed value of $s$.  Observe also that $V_a$ takes
negative values.

Now we can apply electric circuit analysis, in particular applying Kirchhoff's and Ohm's law, to derive
equations respectively describing the adopted simplified common emitter circuit.  For simplicity's
sake, we first consider a purely resistive load $R$, which yields the follwing \emph{time-independent}
equations:

\begin{eqnarray}
   I \left( I_B \right) = \frac{V_{CC} - V_a }{R + Ro(I_B)} \label{eq:IClin} \\
   V_p \left( I_B \right) =  R  \frac{V_{CC} - V_a }{R + Ro(I_B)}  \label{eq:VLlin}  \\
   V_C\left( I_B \right) = V_{CC} -  R \frac{V_{CC} - V_a }{R + Ro(I_B)} \label{eq:VClin} 
\end{eqnarray}

The obtained equations are similar in complexity~\cite{costaequiv:2018} to those obtained for more traditional 
models based on the current gain  $\beta$ and output resistance $R_o = 1/tan(s I_B)$.

However, a slightly more sophisticate situation arises when we considered reactive loads.
Now, because the transistor is non-linear, so that the voltage and current values will vary in a non-linear way
as functions of $f$, we cannot use the customary steady state analysis and phasor methods.  Instead, we need to 
substitute the  reactive components by their integral or differential formulation, so as to define respective 
frequency-dependent voltage or current terms in the respective \emph{time-dependent} circuit equations.  

\begin{figure}[h!]
\centering{
\includegraphics[width=7cm]{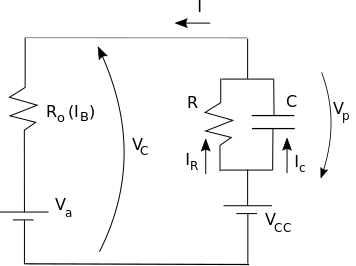}
\caption{The output mesh of the circuit configuration in Figure~\ref{fig:equiv}, with the load
$Z_*$ corresponding to a capacitor $C$ in parallel with a resistor $R$.}
\label{fig:equivpar}}
\end{figure}

We illustrate this procedure with respect to a capacitive load involving a capacitor $C$ in parallel with a
resistance $R$, as illustrated in Figure~\ref{fig:equivpar}.

For this configuration it is convenient to consider the voltage formulation for the load, so that
we have the following, relatively simple, time-dependent equations:

\begin{eqnarray}
   I \left( I_B \right) = \frac{V_{CC} - V_a - V_p(I_B)}{Ro(I_B)} \label{eq:IC} \\
   V_C\left( I_B \right) = V_a + R_o(I_B) \frac{V_{CC} - V_a - V_p(I_B)}{Ro(I_B)} \label{eq:VC} \\
   V_p \left( I_B \right) =  V_{CC} - V_C  \label{eq:VL}  
\end{eqnarray}

At the mesh defined by the load, Kirchhoff's voltage law yields:

\begin{equation}
   V_R I_R = V_C
\end{equation}

At that same mesh, Kirchhoff's current law implies that $I = I_R + I_C$, so that we obtain:

\begin{equation}
   I_C = I - V_C/R_L
\end{equation}

Now we can derive the voltage across the capacitor as:

\begin{equation}
   V_p  =  \frac{1}{C} \int_0^t I_c(t) dt
\end{equation}

The so obtained time-independent and time-dependent equations provide all the necessary resources for analyzing the 
amplification properties of the circuit in Figure~\ref{fig:commemitt} while driving purely resistive and resisitive-reactive
loads.  However, first we provide some considerations regarding the numerical solution of the above circuit.

\section{A Simple Numeric Approach}

Because steady state phasor methods cannot be used to perform AC studies on non-linear circuits with reactive loads,
it is necessary to consider other methods.  In this work, capacitors or inductors are substituted, as implied by each
specific circuit topology, by their respective integro and differential equations, i.e.:

\begin{eqnarray}
   v_C(t) = \frac{1}{C} \int_{t_i}^{t_f} i_C(t) dt \label{eq:Cint} \\
   i_C(t) = C \frac{d v_C(t)}{dt} \label{eq:Cdif} \\ 
   v_L(t) = L \frac{d i_L(t)}{dt}  \label{eq:Ldif} \\ 
   i_L(t) = \frac{1}{L} \int_{t_i}^{t_f} v_L(t) dt \label{eq:Lint} 
\end{eqnarray}

These equations can then be substituted for respective voltages or currents in the circuit, whenever implied by the circuit 
topology and focus of study.  The integrals or differentials are then interactively estimated during the circuit analysis . Let's
illustrate this approach with respect to the circuit in Figure~\ref{fig:equivpar}.   A voltage-based analysis of this
circuit yielded Equations~\ref{eq:IC} to~\ref{eq:VL}.  Assuming that the capacitor is initially discharged (i.e.~$v_C(t=0) = 0V$,
we can derive the following interactive loop, aimed at obtaining the circuit solution from $t_i=0$ to $t$, with time
resolution $dt = t/(Nt-1)$:

\begin{algorithm}[H]
   \caption{Numeric solution with capacitive load.}\label{euclid}
   \begin{algorithmic}[1]
      \Procedure{SolveCap(\textit{Va,s,f,VCC,R,C,Nt,Ib})}{}
         \State $\textit{Vp} \gets 0$
         \For{\texttt{it = 1:Nt}}
            \State $Ro \gets 1 / tan(s*Ib[it])$
            \State $I[it] \gets (VCC - Va - Vp) / Ro$
            \State $VC[it] \gets VCC - Vp$
            \State $Ic \gets IC[it]  -  Vp/R$
            \State $Vp \gets Vp + Ic / C * dt$                                    
         \EndFor
         \State $return(I,VC)$
      \EndProcedure
   \end{algorithmic}
\end{algorithm}

Observe that this algorithm uses a simple trapezoidal integration scheme, so that more demanding situations
may require more precise/sophisticated numeric integration methods (e.g.~\cite{press:2007}).  In the present work
we have used $Nt > 100 000$, but this choice needs to take into account the frequency limit and reactance
and resistance maximum values, so that larger values o $Nt$ may be required in other parametric configurations.

\section{Amplification onto Purely Resistive Loads}

Now we proceed to investigate the characteristics of amplification of the circuit in Figure~\ref{fig:commemitt}
by considering the case of purely resistive loads.  The behavior of the circuit for this configuration is 
given by the time-independent Equations~\ref{eq:IClin} to~\ref{eq:VClin}.

\begin{figure}[h!]
\centering{
\includegraphics[width=8.5cm]{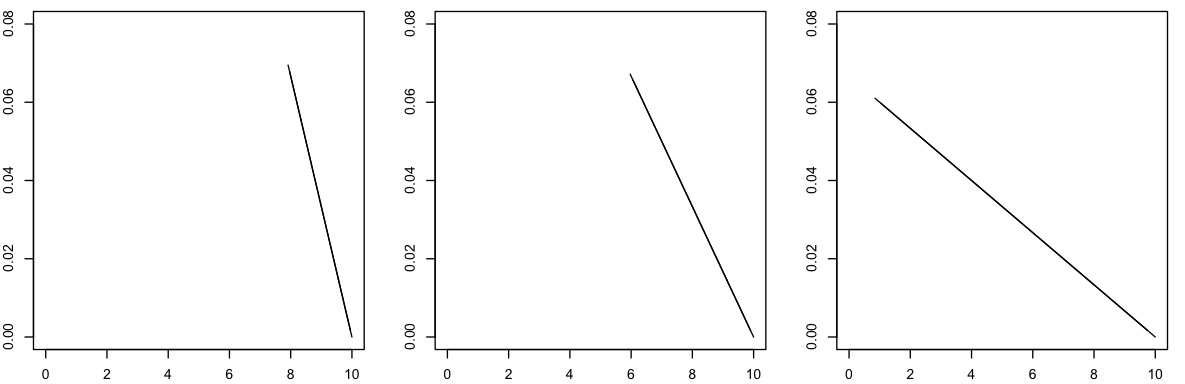}
\caption{The trajectories of the signals $I(t)$  and $V_C(t)$ defined in the $(V_C,I)$ space for purely resistive loads 
$R = 30 \Omega$ (a),  $60 \Omega$ (b)  and $150 \Omega$ (c) are perfectly straight. These results consider $V_a = -50V$
and $s = 10$, a parameter configuration that can be found in silicon PNP BJTs~\cite{costaearly:2018}.}
\label{fig:pureR}}
\end{figure}

We adopt a sinusoidal signal $I_b(t)$ with frequency $1 kHz$ and amplitude $60 \mu A$.    Figure~\ref{fig:pureR} shows 
the trajectories defined by the signals $I(t)$  and $V_C(t)$ in the circuit operation space $(V_C,I)$ for
purely resistive loads with values $R = 30, 60 and 150 \Omega$.  The adopted parameters defining the transistor
operation in these cases were $V_a = -50V$ and $s = 10 V^{-1}$.  These Early parameters values can be found in PNP 
BJT transistors~\cite{costaearly:2018}.

Any trajectory for purely resistive loads $R$ will be, necessarily, fixed and straight.   This is a consequence of the fact that, for this 
type of load, the circuit operation is restricted to the straight line extending from $(V_C = 0, I = V_{CC}/R)$ to $(V_C = V_{CC}, I = 0)$.
Nevertheless, the obtained trajectories will have inclination equal to $1/R$, as in Figure~\ref{fig:pureR}.  So, larger values of $R$
will imply smaller transfer function slopes.

\begin{figure}[h!]
\centering{
\includegraphics[width=8.5cm]{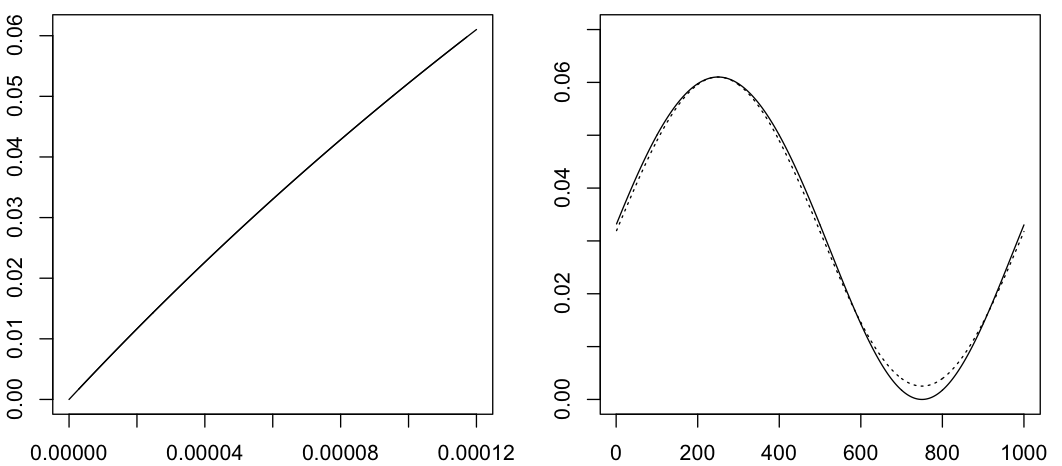}
\caption{The transfer function (a) defined by the considered circuit for $V_a=-50V$ and $s=10$. This curve exhibits a marked
curvature, which implies the amplified current $I(t)$ (b, solid line) to be substantially deformed as compared to a scaled version 
of the pure sinusoidal input $I_B(t)$ (b, dotted line).}
\label{fig:comp_pure}}
\end{figure}

Figure~\ref{fig:comp_pure}(a) shows the transfer function~\cite{lin:2017} defined by the transistor while going from $I_b(t)$ to $I(t)$ 
(i.e.~$I(t) = g(I_B(t))$, and Figure~\ref{fig:comp_pure} (b) shows the superimposition of $I(t)$ and $I_B(t)$, the latter 
having been scaled to match the amplitude of $I(t)$.   It is evident from this figure that the fanned structure of real-world
transistors can strongly deform the original signal, as reflected in the bent transfer function and the mismatch between the shapes
of the input and output currents.  An overall current gain $\langle \beta \rangle \approx s \left| V_a \right| = 500$ was obtained.

\begin{figure}[h!]
\centering{
\includegraphics[width=8.5cm]{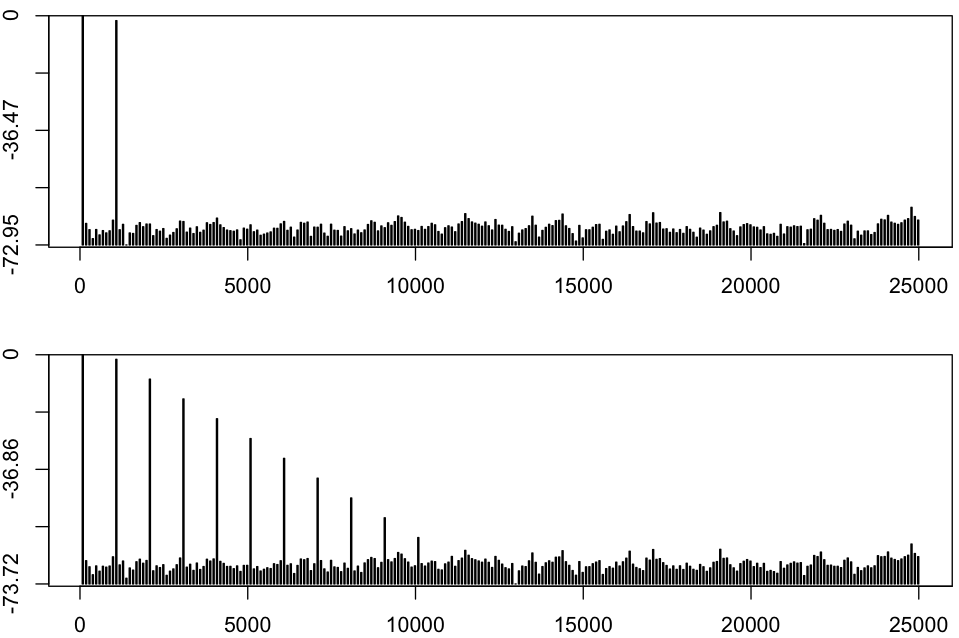}
\vspace{0.5cm} \\
\includegraphics[width=8.5cm]{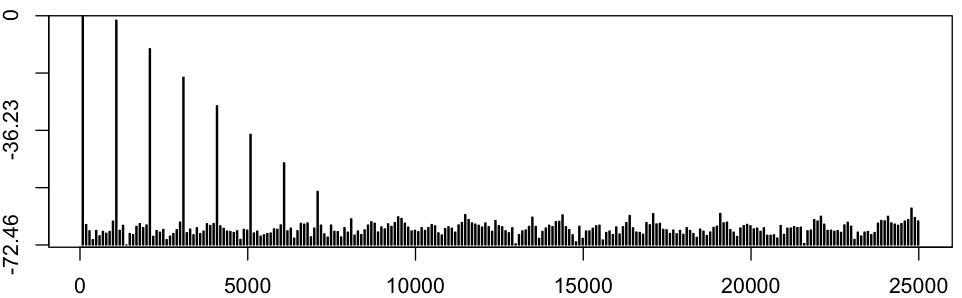}
\caption{The power spectrum for the original pure sinusoidal signal (a) and for the amplified signal $IC(t)$ with $V_a = -50V$ and
$s = 10$.  The power spectrum obtained for $V_a = -200V$ and $s = 2.5$ (same gain) is shown in (b). The smallest
magnitude harmonic components are a consequence of numerical round-off noise.}
\label{fig:power}}
\end{figure}

Figure~\ref{fig:power} depicts the power spectrum of the original signal $I_B(t)$ (a) and of the amplified signal
$I(t)$ (b).  The distortion implied by the transistor geometry on the lower harmonics is evident.  Observe 
that the residual power spectrum values smaller than $-50dB$ are a consequence of numerical round-off noise,
and can be disregarded.

\begin{figure}[h!]
\centering{
\includegraphics[width=8.5cm]{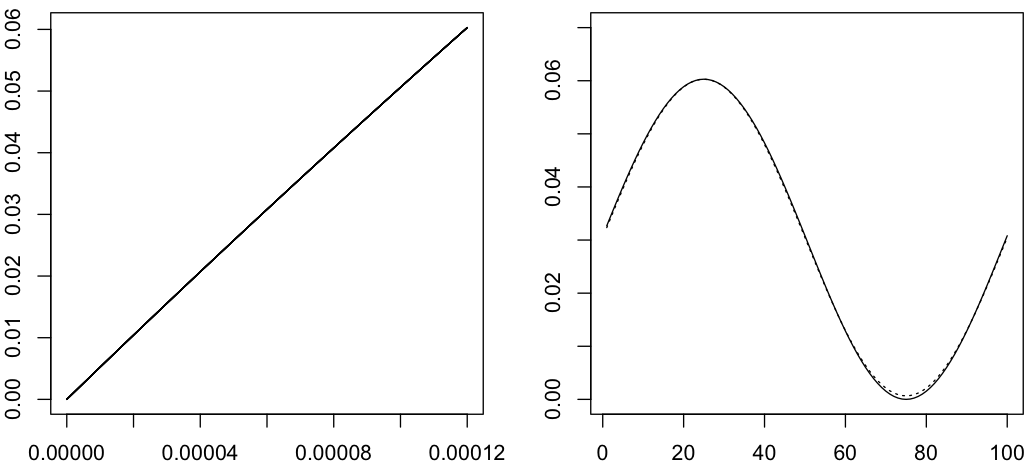}
\caption{The transfer function (a) defined by the considered circuit for $V_a=-200V$ and $s=2.5$. This curve exhibits a less marked
curvature, and hence smaller distortion, than that in Figure~\ref{fig:comp_pure}.}
\label{fig:comp_pure_accur}}
\end{figure}

Now, we repeat the study above, but considering $V_a = -200$ and $s = 2.5$, which yields approximately the same current
amplification.  This parametric configuration can be found in silicon NPN BJTs~\cite{costaearly:2018}.  
Figure~\ref{fig:comp_pure_accur} depicts the respectively obtained transfer function (a) and superimposition of the input
and output currents (b).

The obtained transfer function resulted substantially less bent for for $V_a = -200V$ and $s = 2.5$, implying in much smaller 
amplification distortion.  This confirms that the Early transistor parameters can have great influence on the amplification linearity. 
The power spectrum of the output current  $I(t)$ obtained  is shown in Figure~\ref{fig:power}(c).

The total harmonic distortion (THD) can be employed to effectively summarize the degree of distortion implied by the 
amplification non-linearities, which incorporate new harmonics into the amplified signal.  This index is defined as:

\begin{equation}
    THD = \frac{\sqrt{V_2^2 + V_3^2 + \ldots} }{V_1}
\end{equation}

where $V_i$ is the RMS value of the $i$-th harmonic component, and $V_1$ is the RMS value of the fundamental which, in our
case, corresponds to the pure sinusoidal input $I_b(t)$.  

The THD, considering only harmonic components with relative intensity larger than $1e^{-9}$, obtained for 
$V_a = -50V$ and $s=10$  was found to be equal to $THD(-50,10) = 0.2999995$.   The THD obtained for the
Early parameter configuration $V_a = -200V$ and $s=2.5$ was $THD(-200,2.5) = 0.1499999$, yielding the ratio
$THD(-50,10) / THD(-200,2.5) \approx 2$, which is substantial.   This result corroborates the fact that the fanned
geometric structure underlying real-world transistors can have a major non-linear effect, implying substantial
distortion on the amplification.   Indeed, this could have been already guessed even by visually comparing
the transfer functions obtained for the two considered parameter configurations, as that for $V_a = -50V$ and $s=10$
(Fig.~\ref{fig:comp_pure}) is much more bent than that for $V_a = -200V$ and $s=2.5$ (Fig.~\ref{fig:comp_pure_accur}).

\begin{figure}[h!]
\centering{
\includegraphics[width=8.5cm]{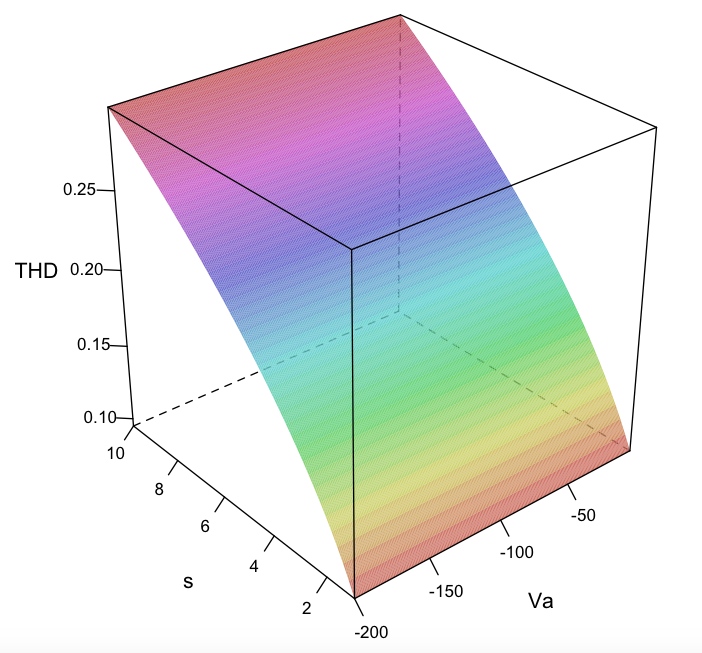}
\caption{The THD values estimated for each of the Early parameter configurations in the rectangular domain $0 \leq s \leq 10$
and $-200V \leq V_a < 0$.  The THD is found to vary only with $s$, and not with $V_a$, increasing in an almost linear way
with $s$.  The THD is also a function of $R$, assuming similar shapes but smaller absolute values for smaller values of $R$.}
\label{fig:THD_pure}}
\end{figure}

Such a strong effect of the transistor Early parameters $V_a$ and $s$ on amplification linearity and distortion motivates a more
systematic analysis.  Figure~\ref{fig:THD_pure} shows the THD values obtained for $R = 150 \Omega$, $V_{CC} = 10V$, and
amplitude of $I_B(t)$ equal to $60 \mu A$.  These values were estimated by using the derived time-varying equations 
to obtain the output signals, followed by respective fast Fourier numerical analysis (e.g.~\cite{brigham:1988,shapebook}).

Remarkably, the distortion implied by the transistor inherent non-linearity \emph{depends only of the transistor Early parameter $s$, not of
$V_a$}. This result is in full agreement with previous Early model investigations of the considered simplified common emitter circuit 
configuration~\cite{costaearly:2018, costaequiv:2018}.  The smaller the value of $s$ of a given transistor, the smaller the THD
will be, and this is the only effect determining distortion as far as the fanning geometrical organization of the transistor isolines is
concerned.  However, considering nearly constant current gains $\beta$, if $s$ is reduced, $V_a$ tends to increase.  So, linearity
can be optimized under the current assumptions by choosing $s$ small and $V_a$ large, which is a characteristic of silicon NPN
BJTs (see~\cite{costaearly:2018, costaequiv:2018}).  Recall that no distortions at all would have been otherwise observed
in case a more traditional model based on current gain $\beta$ and output
resistance $R_o$, with parallel (even though inclined) characteristic isolines had been used.

It can be verified that smaller values of $R$ will yield THD surfaces with similar shapes, 
but with proportionally smaller values.  This is a consequence of the fact that, for smaller values of $R$, the load lines will approach
the vertical, where the current gains implied by the radiating $I_B$-indexed isolines are more uniform.  This interesting
phenomenon is further investigated in the following section.

\section{Amplification onto Capacitive Loads} \label{sec:capload}

The so far obtained results for purely resistive loads corroborate the fact that the non-linearities implied by the fanning structure
of the transistor isolines can impinge large distortions on the respective amplification, depending on the Early parameters
of the chosen transistors and the load resistivity.  Now, we proceed to investigate how this influence
takes place when the load incorporates a capacitive element, as in  Figure~\ref{fig:equivpar}.  

\begin{figure*}[h!]
\centering{
\includegraphics[width=18cm]{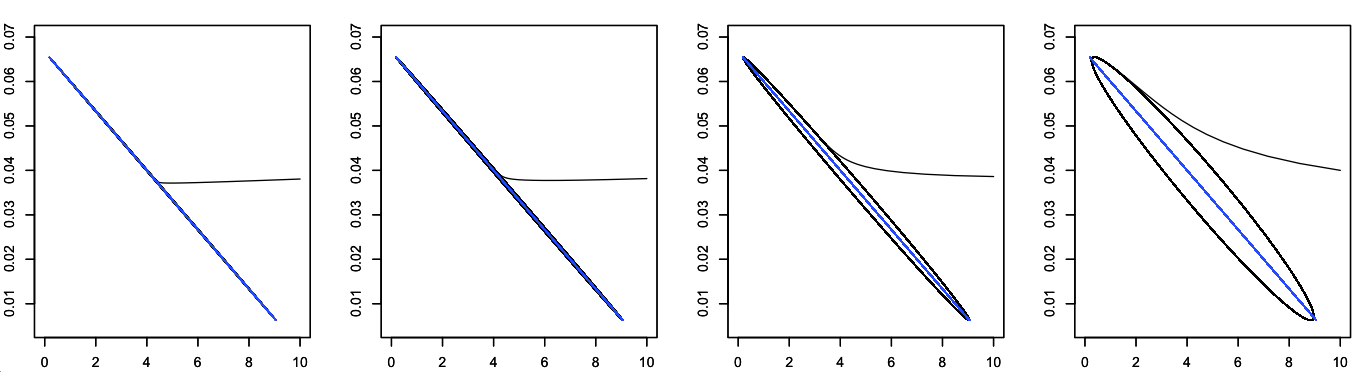}
\caption{The trajectories in the circuit operation space $(V_C,I)$ defined by the input $I_B(t)$ and output $I(t)$ currents for 
$R = 150 \Omega$, $C = 250 nF$ and frequencies $f= 20Hz$ (a),  $f= 70Hz$ (b),  $f= 300Hz$ (c), and  $f= 1 kHz$ (d).
Observe the transient converging (for this range of frequencies) swiftly to an ``ellipsoidal''-like loop.  The detachment of the
trajectory from the resistive load line increases with $f$ to the point (not shown in this figure) as to reduce the $V_C(t)$ voltage gain.
The load lines for $R= 150 \Omega$ are also shown (in blue) for reference, and the considered transistor is characterized by
$V_a = -50V$ and $s=10$.}
\label{fig:load_cap}}
\end{figure*}

First, let's consider the effect of a capacitor $C=250 nF$ placed in parallel with a resistive load $R=150 \Omega$.  Figure~\ref{fig:load_cap}
illustrates the obtained trajectories defined by $V_C(t)$ and $I(t)$ in the circuit operation space $(V_C,I)$ for a pure sinusoidal
inputs  $I_B(t)$ with amplitude $60 \mu A$ and  frequencies $f= 20Hz$ (a),  $f= 70Hz$ (b),  $f= 300Hz$ (c), and  $f= 1 kHz$ (d).
This figure also shows (in blue), as a reference, the load line obtained for a purely resistive load $R = 150 \Omega$.  The
adopted transistor parameters are $V_a = -50V$ and $s=10$.

The transient dynamics can be observed, proceeding from the initial condition to the steady state regime.
As the input frequency increases, the circuit operation progressively detaches from the straight line constraint implied by a pure 
resistive load, defining ``ellipsoidal''-like trajectories.  This is a direct consequence of the fact that, in loads defined by resistance
in parallel with capacitance, the voltage will be ``behind'' (or ``lag'') the current.  However, the non-linearities implied by 
the fanned transistor isolines preclude the application of phasor analysis typically adopted in linear steady state conditions 
(e.g.~\cite{jaeger:1997,hyat:1962}).

\begin{figure*}[h!]
\centering{
\includegraphics[width=18cm]{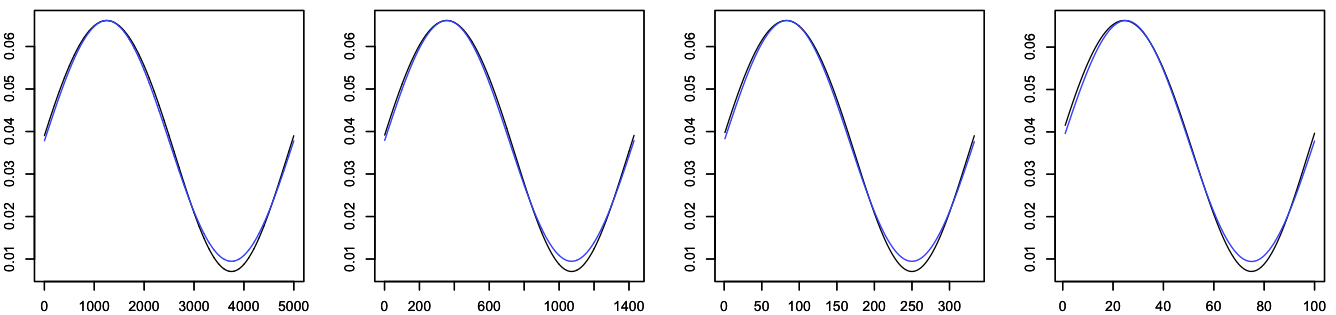}
\includegraphics[width=18cm]{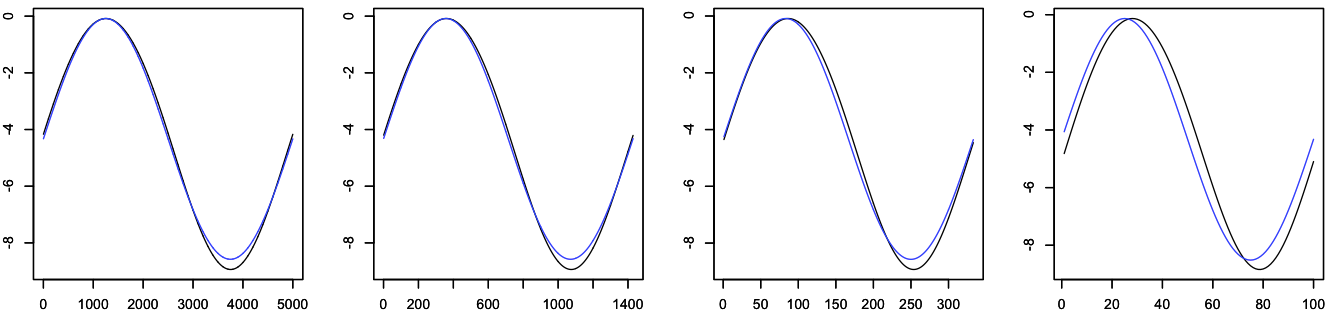}
\includegraphics[width=18cm]{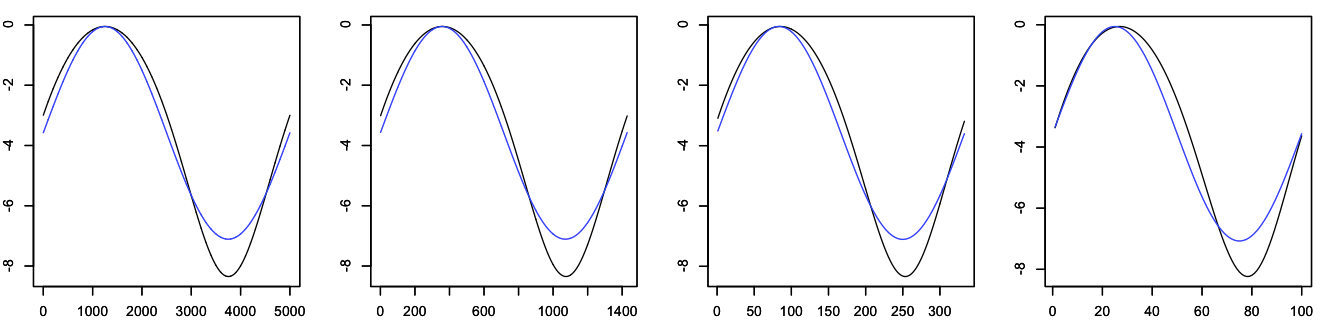}
\caption{The input $I_B(t)$ and output $I_C(t)$ (blue) currents superimposed for comparison of their shapes assuming
$V_a = -50V$, $s=10$, $R = 150 \Omega$ and $V_{CC} = 10V$, respectively to $f= 20Hz$ (a),  $f= 70Hz$ (b),  $f= 300Hz$ 
(c), and  $f= 1 kHz$ (d).  $I_B(t)$ has been suitably
scaled up to match as well as possible (in amplitude) the output current $I_C(t)$.  Observe that only one cycle of the signals
are shown (the time axis is reduced accordingly).  The superimposition of the input current $I_B(t)$ and the output voltage
$V_C(t)$ for $V_a = -50V$ and $s=10$ is also shown in (e)-(h) respectively to $f= 20Hz$ (a),  $f= 70Hz$ (b),  $f= 300Hz$ (c), 
and  $f= 1 kHz$ (d).  The same type of superimposition, but now considering $V_a = -10V$ and $s = 50$ is also shown in
(i)-(l) for the same frequencies. All cycles in this figure were taken after the 3rd loop of the circuit dynamics,
when it was well into the steady state regime. }
\label{fig:comp_cap}}
\end{figure*}

Observe that this detachment of the trajectories  implies the distortions experienced along the upper
trajectory cycle arc to be different from those along the lower arc.  Now, it is a particularly intriguing question to investigate
how this asymmetry can further impact the amplification non-linearity.  Figure~\ref{fig:comp_cap}(a)-(d) shows the input $I_B(t)$ and
output signals $I(t)$ obtained for each of the four cases in Figure~\ref{fig:load_cap} ($V_a = -50V$ and $s=10$).  
Observe that $I_B(t)$ has been suitably scaled up so that its shape can be directly compared to that of $I_C(t)$.

\begin{figure}[h!]
\centering{
\includegraphics[width=7cm]{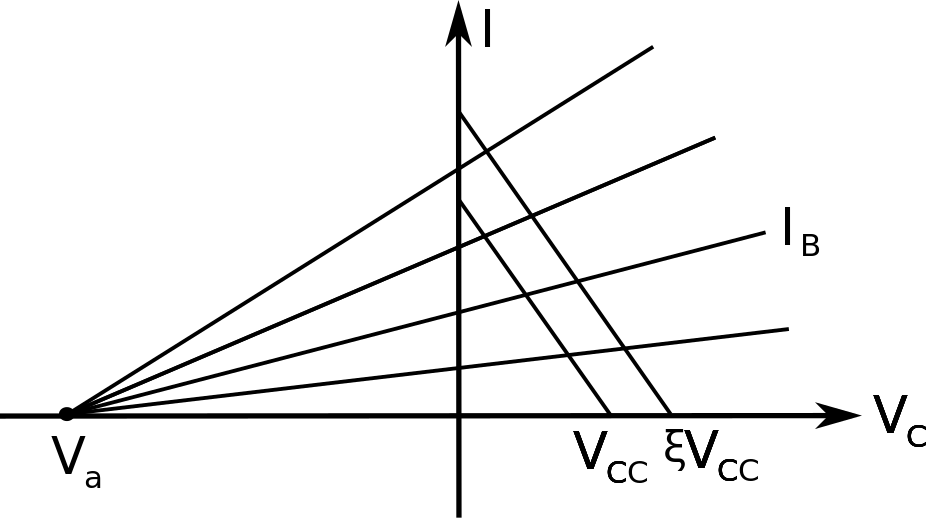}
\caption{The geometrical construction that can be adopted in an approximate study of the asymmetry of gains in the upper and lower
arcs of capacitive loops in load lines.  The two arcs can be approximated by two parallel straight load lines, one for $V_{CC}$ (lower arc
of the loop) and the other for an incremental increase $\xi V_{CC}$ (upper arc).  Remarkably, as derived in the text, the ratio between 
the current gains along any such two parallel (same $R$) load lines, even when further apart, does not depend on $s$ or $R$, being 
solely a function of $V_{CC}$, $V_a$ and $\xi$.}
\label{fig:increm}}
\end{figure}

Interestingly, the distortions implied by this circuit configuration are very similar to those observed in Figure~\ref{fig:comp_pure}(b).
Indeed, THD comparisons showed that the trajectory asymmetry implied by the capacitive load does not affect significantly --
at least for the considered circuit, parameter and variable configurations -- the linearity of the amplification.   This is an interesting
phenomenon that can be further investigated analytically by using two parallel purely resistive load lines, as in Figure~\ref{fig:increm} 
to approximate the gain variation as the trajectory excursions along the upper and lower arcs of the loop.  Observe that the 
displacement of these load lines is defined by the supply voltage $V_{CC}$.

We have from the Early modeling of the considered circuit with a purely resistive load $R$ has its output current $I(I_B)$ expressed as:

\begin{equation}
  I(I_B) = \frac{V_{CC} - V_a}{R + R_o(I_B)} = \frac{\left( V_{CC} - V_a \right) \, tan (s I_B)}{R \, tan(s I_B) + 1}
\end{equation}

It follows that the current gain $\beta(I_B)$ along the load line, in terms of $I_B$ is given as follows:

\begin{equation}
  \beta(I_B) = \frac{\partial{ I(I_B)}}{\partial{I_B}}  = \frac{s \left(  V_{CC} - V_a \right) csc^2(s \, I_B)}{ \left( cot(s \, I_B) + R \right) ^2}
\end{equation}

Observe that this current gain is a function of the transistor Early parameters $V_a$ and $s$ as well as the circuit parameters
$R$, and $V_{CC}$, as well as $I_B$.  Now, we take the output current $I_{\alpha}(I_B)$ along a load line parallel (same $R$) to the
previous one.   This new load line has it position along the horizontal axis determined by $\xi V_{CC}$, where $\xi$ is a 
proportionality coefficient (e.g.~$\xi = 1.1$):

\begin{equation}
  I_{\xi}(I_B) =  \frac{\left( \xi V_{CC} - V_a \right) \, tan (s I_B)}{R \, tan(s I_B) + 1}
\end{equation}

The current gain $\beta_{\alpha}(I_B)$ along this load line, in terms of $I_B$, can be derived as:

\begin{equation}
  \beta_{\xi}(I_B) =  \frac{s \left(  \xi V_{CC} - V_a \right) csc^2(s \, I_B)}{ \left( cot(s \, I_B) + R \right) ^2}
\end{equation}

Now we can obtain the proportion $\rho$ between the current gain between the second and first load lines as:

\begin{equation}
  \rho = \frac{\beta_{\xi}(I_B)} {\beta(I_B)} =  \frac{\xi V_{CC} - V_a}{ V_{CC} - V_a}
\end{equation}

\begin{figure}[h!]
\centering{
\includegraphics[width=8cm]{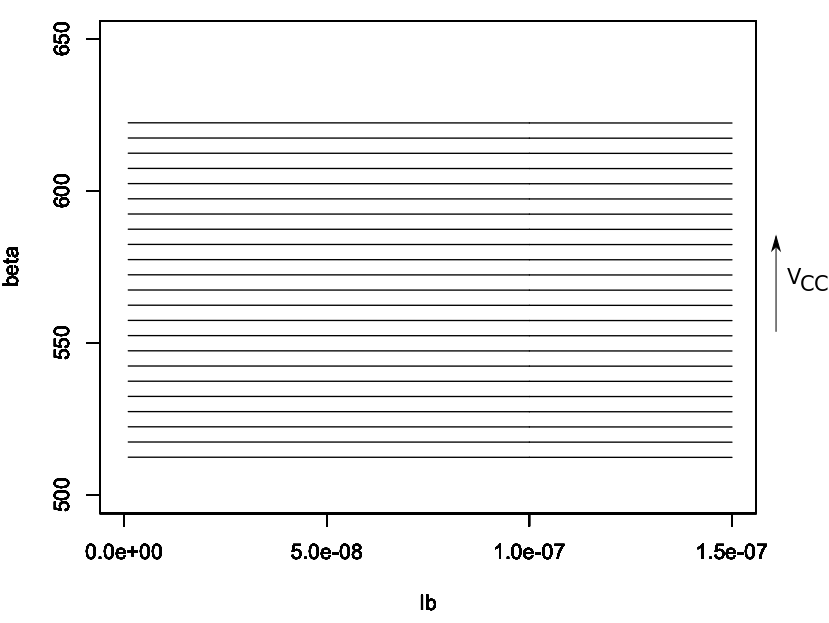}
\caption{$\beta$ is constant with $I_B$ for $5 \leq V_{CC} \leq 50$, increases with equispaced
steps as the $V_a$ magnitude is drecreased.  Results obtained for $R = 150 \Omega$, $V_a = -200$ and $s=2.5$
with $50V$ steps.}
\label{fig:beta_f_VCC}}
\end{figure}

Remarkably, $\rho$ turns out \emph{to be independent} of $I_B$, $s$ and $R$, varying only with $V_{CC}$, $V_a$ and $\xi$.  
This is shown in Figure~\ref{fig:beta_f_VCC} for $5 \leq V_{CC} \leq 50$, where successive constant gains are obtained for
$V_{CC}$ steps of $50V$.

\begin{figure}[h!]
\centering{
\includegraphics[width=8cm]{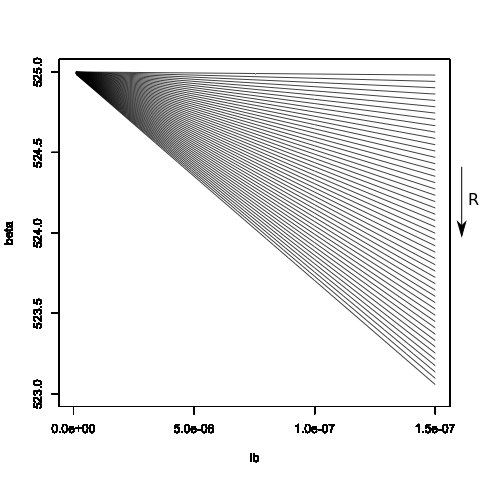}
\caption{The current gain $\beta(I_B)$ decreases with $I_B$.  The rate of decrease is larger for larger values of $R$.
The gain variations are minimized when $R$ is small, achieving perfectty linear amplification in the limit $R \rightarrow 0$.}
\label{fig:beta_f_Ib}}
\end{figure}

Figure~\ref{fig:beta_f_Ib} illustrates the current gain $\beta(I_B)$ in terms of $0 \leq I_B \leq 15 \mu A$, for 
$50 \Omega \leq R \leq 5 k \Omega$.  As a consequence of the $I_B$-indexed fanned characteristic of the isolines underlying
the transistor operation, the current gain varies in a virtually linearly decreasing way with $I_B$ for all considered 
values of $R$.  Remarkably, as less gain variation implies in enhanced linearity, \emph{almost perfect linear amplification is
achieved for loads with very smal} $R$.  Indeed, perfect linear operation is achieved in the limit $R=0 \Omega$, provided $I_B(t)$
is controlled so as to have $I_C(t)$ not exceeding the voltage supply maximum current and the transistor maximum power dissipation.

Now, substituting for $V_a = -200V$, $V_{CC} = 10V$ and $\xi = 1.1$, we have that $\rho = 1.004762$.  For a good average gain (NPN) 
$\beta = s V_a = 2.5 |-200| = 500$, this implies in a relative gain variation of $((500) 1.004762) +500)/500 = 0.004762$, which is 
negligible given the typically obtained values of gains, parameters and variables for the considered configurations.

So, for the considered hypothesis and configurations, the capacitive component of the load does not strongly influences the
linearity of the amplification despite the respectively implied asymmetry along the upper and lower arcs of the loop.  This happens
particularly for large values of $V_a$, which is known~\cite{costaearly:2017,costaearly:2018,costaequiv:2018} 
to be directly related to the transistor average output resistance
$\langle R_o(I_B) \rangle$, irrespectively of $s$.  So, transistors with larger values of $V_a$ not only will have better linearity
for a same gain, but will also (and for the same reasons) imply better resilience to capacitive load effects.  However, large
output resistance also implies bigger lag between the output voltage and current, and hence typically unwanted low-pass filtering,
which is another important aspect deserving further consideration.

Figures~\ref{fig:comp_cap}(e)-(h) show the collector voltage $V_C(t)$ superimposed onto the input current $I_B(t)$,
with respect to $f= 20Hz$ (e),  $f= 70Hz$ (f),  $f= 300Hz$ (g), and  $f= 1 kHz$ (h), assuming $V_a = -50V$ and $s=10$.   Observe
that $V_C$ had to have its signal inverted, and then $I_B(t)$ was scaled up so as to allow the comparison between the shape
of these two signals.  Unlike what was verified between the input and output currents, which is always in phase, there is
a very perceptible lag between the output voltage and current that increases with $f$, even though this is not a particularly
high frequency. So, the choice of large values of $V_a$ implies larger average output resistance $\langle R_o \rangle$ that 
tends to worsen the lag, implying more intense low-pass filtering.  So, a trade-off is revealed between linearity and
low-pass filtering implied by capacitive loads.  This tends to become particularly critical for loads with large resistive or
capacitive components (i.e. large $\tau = RC$).   

Because filtering represents a distortion of the original signal, it is important to try to devise means for avoiding the effects
of capacitive loads.   An immediate possibility is to use transistors with smaller $V_a$, leading to smaller average output
resistance $\langle R_o \rangle$.  Figures~\ref{fig:comp_cap}(i)-(l) depict the output voltage $V_C(t)$ and input current
$I_B(t)$ for $V_a = -10V$ and $s=50$ (same gain as before, but much smaller output resistance), with respect to
 $f= 20Hz$ (e),  $f= 70Hz$ (f),  $f= 300Hz$ (g), and  $f= 1 kHz$ (h).  Though a large distortion has been implied to the
 voltage signal, the smaller output resistance allowed by $V_a = -10V$ allowed an almost perfect phase match between
 the input current $I_B(t)$ and the output voltage $V_C(t)$.  However, the severity of the impinged distortion renders this
 solution unviable for most applications, except as a possible means of dealing with synchronization problems.

Another possibility to avoid the complications implied by capacitive loads is to add an inductor in series with the load,
with reactance perfectly matched so as to cancel the load capacitance (in a manner remindful of load correction), 
yielding a purely resistive load as a result.   Interestingly, the compensation required for any given signal frequency 
will also work  for other frequencies.  This possible technique is allowed as a consequence of the transistor non-linearities 
having been found in this work, at least for the considered hypotheses, not to influence too much the asymmetry of 
the amplification along the two arcs of reactive loops, otherwise the compensation could be undermined.  This possible technique 
also requires the capacitance to be canceled not to be a function of other parameters (e.g. voltage, electromagnetic
field or temperature) than the signal frequency.

\section{Concluding Remarks}

Linear amplification remains a critically important issue for a vast number of electronics applications required by
science/technology and as assistance to humans.  However, the
non-linearity intrinsic to transistors has substantially constrained more systematic and complete studies of transistor-based
amplification.  The recent development of a simple and yet accurate transistor model~\cite{costaearly:2017,costaearly:2018,costaequiv:2018} 
capable of incorporating the key non-linearities characterizing the operation of these devices, paved the way to
allowing more accurate studies of transistor amplification.  One particularly relevant and challenging aspect related to the
non-linearity of transistor amplification concerns the influence of loads with reactive components on the circuit
operation.  This issue has been particularly complicated to be approached because of the lack of simple and accurate
transistor models capable of reflecting the fanned structure of the characteristic surface of real-world transistors.
This structure implies non-linearities precluding the complex phasor approach commonly used in AC circuit analysis
to be used.  To address this problem in a more systematic way provided the main motivation for the present work.

A simplified common emitter circuit configuration was adopted and represented in terms of its Early model, which involves
only two parameters, $V_a$ and $s$, that are invariant to the circuit operation as defined by the base current $I_B(t)$,
collector current $I(t)$ and collector voltage $V_C(t)$.  The Early equivalent circuit of a transistor was then applied to derive
simple, and yet accurate time-independent and time-dependent voltage and current equations describing the circuit behavior 
for purely resistive loads as well as for reactive loads.  The approach in this work focused capacitive loads because the 
inductive counterpart can be treated in a completely analogue way except for the sign of relative lag between the output voltage
and current.  As the non-linearities implied by the fanned characteristic of real-world transistors preclude the application of
well-established phasor AC analysis, we resorted to a numeric-interactive solution in which the integral expression for the voltage
in a capacitor is integrated by using a simple numerical method (trapezoid method), which was presented in detail.

In order to provide a reference for the analysis of capacitive loads, we investigated the purely resistive case first.  This 
was done by using the simpler time-independent circuit equations directly derived from the Early approach.  Several interesting
results were obtained.  In particular, it was shown that the Early parameters $V_a$ and $s$ have major effect on the linearity
of the transistor amplification.  This effect is revealed mostly as a saturation of the current gain as $I_B$ increases, as 
a consequence of the transistor radiating isoline characteristics.  This implies a gain asymmetry between the positive and
negative cycles of the AC version of the input current.   Transistors with larger $V_a$ magnitude and smaller $s$ tend to yield 
improved linearity for the same gain of transistors with smaller $V_a$ magnitude and larger $s$ capable of delivering the same
current gain.  Power spectra obtained for such different parameter choices revealed a linearly decaying magnitude
along the lower harmonics, an effect that is much more pronounced in the case of large $s$ and smaller $V_a$.  This
different spectral composition motivated the consideration of total harmonic distortion (THD) measurements.  These measurements
were easily obtained by using the time-dependent version of the circuit equations, followed by fast Fourier transform, allowing
the systematic identification of the THD for each possible parameters configuration in wide region of the Early parameter space $(V_a,s)$.
The obtained, more accurate, results confirmed previous related studies using gain approximation equations~\cite{costaequiv:2018}, 
with the \emph{THD depending only of $s$ in an almost linear way, while being completely independent of} $V_a$.  The THD
for the considered circuit and parameter configurations was also found to decrease with smaller values $R$ of the resistive load,
with the remarkable result that \emph{perfectly linear amplification is achieved for} $R \rightarrow0 \Omega$, provided the transistor and
circuit maximum dissipations are observed.

The reactive load situation was addressed next, with focus on capacitance.  This case is particularly interesting because the lag
between voltage and current, which increases for larger signal frequencies, implies the trajectory defined by the circuit operation in the
$(V_c,I)$ space to detach from the straight line characterizing purely resistive loads and for ``ellipsoidal''-like loops.  So, as the
circuit goes along the upper or lower arcs of this loop, as $I_B$ varies, it experiences different non-linearities 
as a consequence of the fanned geometry of transistor amplification.  This could, in principle, complicate the intrinsic non-linearities
already observed for purely resistive loads.  In the case of typical Early parameter values configurations, and under the considered
choices and parameter configurations, it was found numerically and theoretically that the transistor linearities implied by the loop-induce
gain asymmetry are relatively minor in the sense that most of the impinged distortions are a consequence of the transistor fanned structure
solely on the resistive load.  The analytical approach to this result involved the construction of an approximation to the loop-induced
gain asymmetry by using two parallel straight load lines.  It was remarkably shown that under these circumstances the ratio of the
current gains along the two parallel load lines \emph{does not depend on} $I_B$ or $R$.  As a matter of fact, the current gain was found to
vary in almost perfectly linearly decreasing fashion with $I_B$, which accounts for the gain independence of $I_B$.

As the transistor non-linearities were found not to significantly contribute additional distortions as a consequence of the gain
asymmetries along the upper and lower arcs of the loop, choosing transistors with larger $V_a$ will improve the linearity of the
amplification.   However, this also increases the average output resistance $\langle R_o \rangle$, implying in larger lags between
voltage and current.  A trade-off is therefore established between linearity and phase lag (low-pass filtering) by choice of larger
$V_a$.  The possibility to try reducing the lag by using smaller $V_a$ (and therefore smaller $\langle R_o \rangle$ resulted
in almost null lag, but also in severe signal distortion.  As an alternative for avoiding the unwanted capacitive effects on transistor
amplification, a series inductor can be incorporated to the load which, when properly matched, can eliminate the capacitive
behavior.  This requires both capacitance and inductance not to depend on other parameters such as temperature or voltage.

The reported methodology and results pave the way to many future investigations.  In particular, it would be interesting to
consider other amplifier configurations, such as more conventional common emitter circuits.  In such cases, it would be 
also worth studying the contribution of negative feedback on enhancing linearity, as it has been experimentally shown that
negative feedback may not be able to completely eliminate parameter variations that are intrinsic to real-world transistors.
It would also be interesting to consider other parameter configurations, such as higher frequencies and $V_CC$, as well
as more substantial capacitance and resistor values.  The remarkable lag elimination provided by transistors with small
$V_a$ and large $s$ can also be eventually applied in electronic synchronization problems.  Another research worth of
future investigation is the case of controllable loads, such as amplification taking place over transistors.

\vspace{0.7cm}
\textbf{Acknowledgments.}

Luciano da F. Costa
thanks CNPq (grant no.~307333/2013-2) for sponsorship. This work has benefited from
FAPESP grants 11/50761-2 and 2015/22308-2.
\vspace{1cm}


\bibliography{mybib}
\bibliographystyle{plain}
\end{document}